\documentclass[superscriptaddress,showpacs,aps,twocolumn,floatfix]{revtex4-1}
\usepackage{bm,amsmath,amssymb}
\usepackage{amssymb}
\usepackage{amsmath}
\usepackage{commath}
\usepackage{graphicx,bm}
\usepackage{verbatim}
\usepackage[latin1]{inputenc}
\usepackage{xcolor}
\usepackage[title]{appendix}

\newcommand{\fref}[1]{Fig.~\ref{#1}}

\newcommand{\sref}[1]{Sec.~\ref{#1}}

\begin{document}

\title{Optimal  thermoelectricity with quantum spin-Hall edge states}

\author{Daniel Gresta}
\affiliation{International Center for Advanced Studies, ECyT-UNSAM, Campus Miguelete, 25 de Mayo y Francia, 1650 Buenos Aires, Argentina}
\author{Mariano Real}
\affiliation{Instituto Nacional de Tecnologia Industrial, INTI, Av. General Paz 5445, (1650) Buenos Aires, Argentina}
\author{Liliana Arrachea}
\affiliation{International Center for Advanced Studies, ECyT-UNSAM, Campus Miguelete, 25 de Mayo y Francia, 1650 Buenos Aires, Argentina}

\begin{abstract}
We study the thermoelectric properties of a Kramer's pair of helical edge states of the quantum spin Hall effect coupled to a nanomagnet with a component of the magnetization perpendicular to the direction of the spin-orbit interaction of the host. We show that the transmission 
function of this structure has the desired qualities for optimal thermoelectric performance in the quantum coherent regime. For a single magnetic domain there is a power generation close to the optimal bound.
In a configuration with two magnetic domains with different orientations, pronounced peaks in the transmission functions and resonances lead to a high figure of merit. We provide estimates
for the fabrication of this device with HgTe quantum-well topological  insulators.
\end{abstract}

\maketitle

{\em Introduction.} Thermoelectricity in the quantum regime is attracting high interest for some years now \cite{giazotto,casati}. Systems hosting edge states, like the quantum Hall and quantum spin Hall are paradigmatic realizations of quantum coherent transport. Several theoretical and experimental results on heat transport and thermoelectricity in these systems have been recently reported \cite{granger,Nam,us,qcond,cappelli,grosfeld,us1,stern,heiblum,altimiras,yacoby,altimiras2,baner,half,pheno,torsten,rafa,peter,janine,enhan,vanuci,jauho,rone,roda,graph,bjorn,helius,soth,benja}. 
 
Unlike the quantum Hall state, which is generated by a strong magnetic field, the  quantum spin Hall (QSH) state taking place in two-dimensional (2D) topological insulators (TI), preserves time-reversal invariance. Therefore, the edge states appear in helical Kramer's pairs  \cite{ti1,ti2,ti3,ti4,ti5,ti6} with opposite spin orientations determined by the spin orbit of the TI.
Several heat engines and refrigerators  have been recently proposed, taking advantage of the fundamental chiral nature of the quantum Hall edge states, which manifests itself in multiple-terminal structures\cite{rafa} and in quantum interference\cite{peter,janine}. Recently, the  property of charge fractionalization was also pointed out as a mechanism to enhance thermoelectricity \cite{enhan}. All these setups rely on the existence of quantum point contacts and quantum dots  in the structure, 
tunnel-coupled to the edge states, which are generated by 
recourse to constrictions. The fabrication of these elements is nowadays normal in the context of the quantum Hall effect \cite{qpc1,qpc2,qpc3}. 
However,  their realization  in the context of the QSH effect remains an experimental challenge so far \cite{qpc4}, although they  are widely investigated theoretically  
\cite{dolcini,citro,rech1,bruno1,tidot1,tidot2,tidot3,tidot4,tidot5,bruno2}. 

\begin{figure}
	\centering
	\includegraphics[width=0.8\columnwidth]{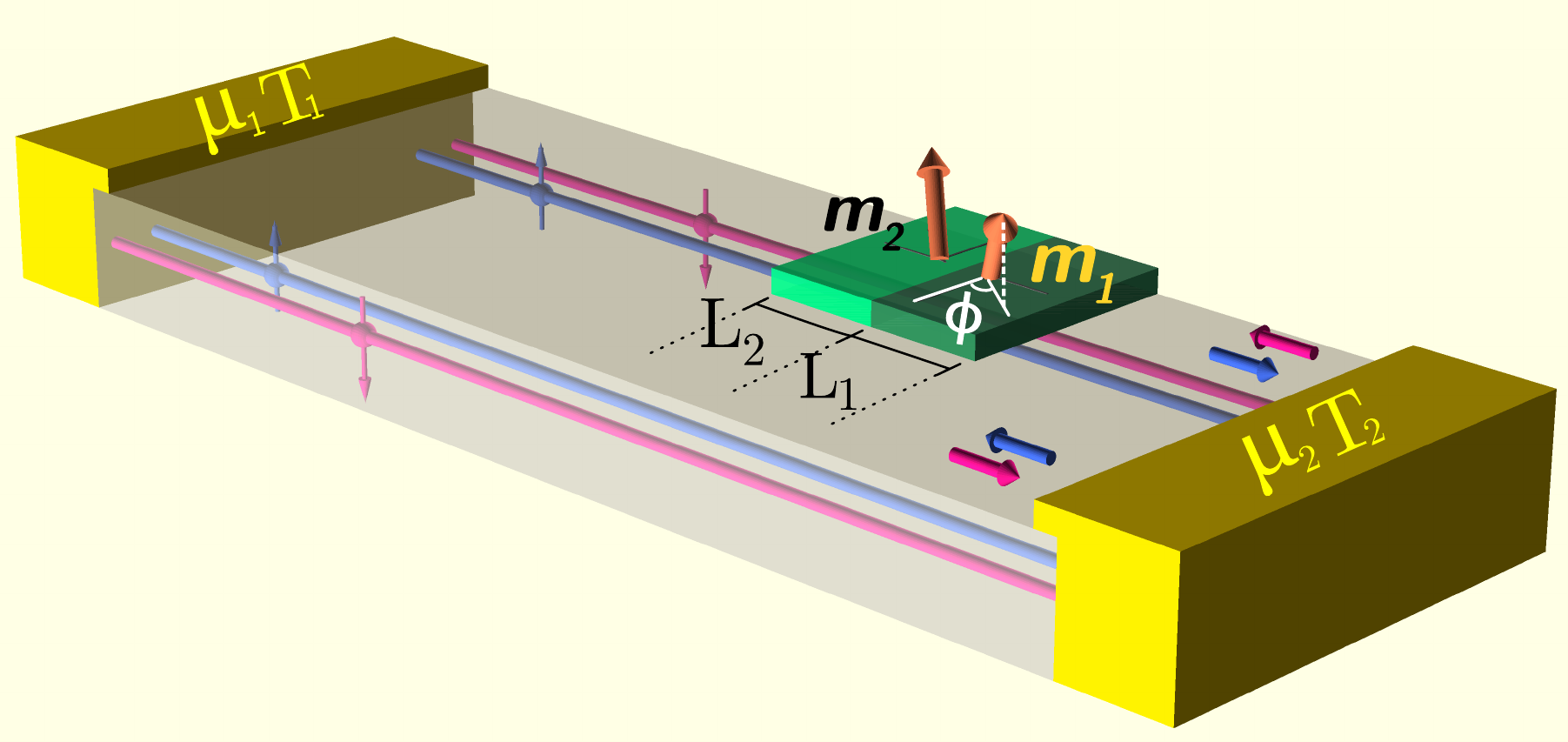}
	\caption{Sketch of the setup scheme. 2D TI  contacted to ohmic contacts at which a bias voltage $eV=\mu_1-\mu_2$ and temperature difference
	$\Delta T=T_2-T_1$ are applied. Two nanomagnets with magnetic moments ${\bf m}_1$ and ${\bf m_2}$ and lengths $L_1$ and $L_2$,
	 are contacted to a helical  Kramer's pair of edge states.}
	\label{fig:fig0}
\end{figure}

In the quantum coherent regime the electronic transport properties take place without inelastic scattering and are fully characterized by a transmission function. Particle-hole symmetry breaking is a necessary condition for steady-state heat to work conversion. Having transmission functions rapidly changing in energy within the relevant transport window, is the key to achieve optimal  thermoelectricity \cite{casati,maso,benenti,whitney1,whitney2}. The optimal performance is usually quantified by the {\em figure of merit} $ZT$, with the  Carnot limit achieved 
for $ZT \rightarrow \infty$. This ideal limit can be realized for transmission functions containing delta-function like peaks \cite{maso}. In this sense, structures with resonant levels like quantum dots are particularly promising \cite{linke,fabio1,fabio2,dutta,ora,arman,misha1}. 
On another hand, electrical {\em power generation} out of heat is the aim of thermoelectric heat engines. This is optimized by transmission functions behaving like Heaviside-theta functions within the relevant transport window \cite{whitney1,whitney2}. In the case of quantum-Hall edge states, configurations with several quantum point contacts and quantum capacitors, have been recently proposed to engineer the transmission function for optimal thermoelectricity by recourse to quantum  interference \cite{peter}. 

In the present work we analyze a very different mechanism for edge-state thermoelectricity in a QSH structure. It is based on the coupling of a Kramers pair of helical edge states of the QSH to a magnetic domain. The structure we analyze is sketched in Fig.\ref{fig:fig0}, where an edge-state pair of a 2D TI is contacted by nanomagnets. We consider two configurations, a single magnetic island with a given magnetic orientation,  as well as two attached islands with different orientations of the magnetic moments. In both configurations, the key ingredient is a finite component of the magnetic moments perpendicular to the direction of the spin-orbit interaction of the TI. A similar structure was previously considered in Refs. \onlinecite{magdot1,magdot2,magdot3}, focusing on the interplay between  spin-torque induced current and the consequent pumping induced by the precession of the magnetic moment.
In combination to superconducting contacts, this structure has been investigated as a platform to realize topological superconductivity \cite{fu,meyer,crepin}. Here, we show that the simple two-terminal setup of Fig.\ref{fig:fig0}, under the effect of a simultaneous voltage and temperature biases, has the desired properties for optimal 
heat to electrical work dc conversion. We analyze the transmission function for this structure and the impact of its different features on the thermoelectric response. Remarkably, this function takes the best of the two worlds regarding power generation and large figure of merit, since it has features alike a theta function and delta-function type resonances due to bound states in the gap, as well as peaks alike  quantum dots. We provide estimates for the different components of the device and we argue that it is within the present state of the art of
fabrication of 2D TI structures \cite{ti4,ti5,ti6}.

{\em Thermoelectric performance in the quantum coherent regime}. We briefly review the linear-response thermoelectric approach assuming small differences of chemical potential $eV=\mu_1-\mu_2$ (with $\mu_1=\mu$), and temperature $\Delta T=T_2-T_1$ (with $T_1=T$), applied at the contacts of the edge states, as indicated in Fig. \ref{fig:fig0} \cite{casati}.
 The induced charge and heat currents  are
\begin{equation}
\left(
\begin{array}{c}
I^C/e \\
I^Q 
\end{array}
\right)  =  \left(
\begin{array}{cc}
{\cal L}_{11} &{\cal  L}_{12}  \\
{\cal L}_{21} &{\cal L}_{22} 
\end{array}
\right) 
\left(
\begin{array}{c}
X_1\\
X_2
\end{array}
\right).
\end{equation}
We have introduced the affinities $X_1=eV /k_B T$ and $X_2= \Delta T/k_B T^2$. In the quantum coherent regime, the elements of the
Onsager matrix are fully determined by the transmission function ${\cal T}(\varepsilon)$ as follows,
\begin{equation}
{\cal L}_{ij} = - T \int \frac{d \varepsilon}{h} \frac{\partial f (\varepsilon)}{\partial \varepsilon} \left(\varepsilon-\mu \right)^{i+j-2} {\cal T}(\varepsilon),
\end{equation}
where  $f(\varepsilon)=1/(e^{(\varepsilon-\mu)/k_B T}+1)$.
The key for the thermoelectric heat to work conversion is encoded in the off-diagonal coefficient 
${\cal L}_{12}={\cal L}_{21}$. 
The quality of this conversion is evaluated in terms of
the efficiency (for the heat engine), $\eta^{\rm he}=(I^C T X_1)/I^Q$, being $P=I^C T X_1$ the generated power,
 or coefficient of performance (for the refrigerator), $\eta^{\rm fri}=-I^Q/I^C T X_1$, being $-I^Q$ the heat current extracted from the cold reservoir. In both cases, for a given difference of temperature, the maximum 
values for these coefficients can be parametrized by the figure of merit $ZT={\cal L}_{21}^2/\mbox{Det}{\hat{\cal L}}$, as follows
\begin{equation}\label{zt}
\eta^{\rm max}= \eta_C^{\rm he/fri} \frac{\sqrt{ZT+1}-1}{\sqrt{ZT+1}+1},
\end{equation}
being $\eta_C^{\rm he}=\left[\eta_C^{\rm fri}\right]^{-1} = \Delta T/T$ the Carnot efficiency, which is achieved for $ZT \rightarrow \infty$, while the value  $\eta^{\rm he/fri} \sim 0.3 \; \eta_C$ corresponds to $ZT \sim 3$. As originally shown
by Mahan and Sofo, the ideal upper bound $\eta_C^{\rm he/fri}$ is obtained for ${\cal T}(\varepsilon) \sim \delta(\varepsilon-\varepsilon_0)$, while $ZT$ attains high values when ${\cal T}(\varepsilon)$ has 
peaks within the relevant transport window $|\varepsilon-\varepsilon_0| \sim k_B T$. 
On another hand, for the heat engine operational mode,
the maximum achievable power for a given $\Delta T$  and the corresponding efficiency are
\begin{equation}\label{pmax}
P_{\rm max}= \eta_C \frac{{\cal L}_{12}^2 X_2}{4{\cal L}_{11}},\;\;\;\;\;\;\eta( P_{\rm max})= \eta_C \frac{ZT}{2(ZT+2)}.
\end{equation}
It has been shown that the maximum power is bounded by $0.32 P_0$ for a transmission function of the form ${\cal T}(\varepsilon) \sim \theta(\varepsilon -\mu-\varepsilon_0)$, where $P_0=(k_B \Delta T)^2/h$ \cite{whitney1}.

{\em Transmission function.} 
The structure sketched in Fig.\ref{fig:fig0} is modeled by the following Hamiltonian
\begin{equation}\label{ham}
H= \int dx \Psi^{\dagger}(x) \left[\left(-i \hbar v_F \partial_x  \right) \hat{\sigma}_z +  J {\bf m}(x) \cdot \hat{\boldsymbol{\sigma}} \right]\Psi(x),
\end{equation}
where $\Psi(x)=\left(\psi_{R,\uparrow}(x), \psi_{L,\downarrow}(x)\right)^T$, represent the right(left)-moving electrons with velocity $v_F$ and $\uparrow$ ($\downarrow$) spin orientations, $J$ is the magnetic exchange interaction between the magnetic moment of the island and the spin of the electrons, and $\hat{\boldsymbol{\sigma}} =\left(\hat{\sigma}_x,\hat{\sigma}_y,\hat{\sigma}_z\right)$ are the Pauli matrices. 
 The magnetic island is described by the following piece-wise spacial distribution 
 of the magnetic moment within segments of  lengths $L_j=x_j-x_{j-1}$. 
\begin{equation}
{\bf m}(x) =  \sum_{j=1}^N \theta(x_j-x) \theta(x-x_{j-1}) {\bf m}_j.
\end{equation}
${\bf m}_j = \left(m_{j \perp} \cos \phi_j, m_{j \perp} \sin \phi_j, m_{j ||}\right)$ is the magnetic moment per unit length with components $m_{j ||}$ (parallel)  and $m_{j \perp}$ (perpendicular)
with respect to the direction of the spin-orbit interaction of the TI. We focus on a single island ($N=1$) and two islands $N=2$ of the same length but with different orientations of the
magnetic moment. 

\begin{figure}
\begin{center}
\includegraphics[width=\columnwidth]{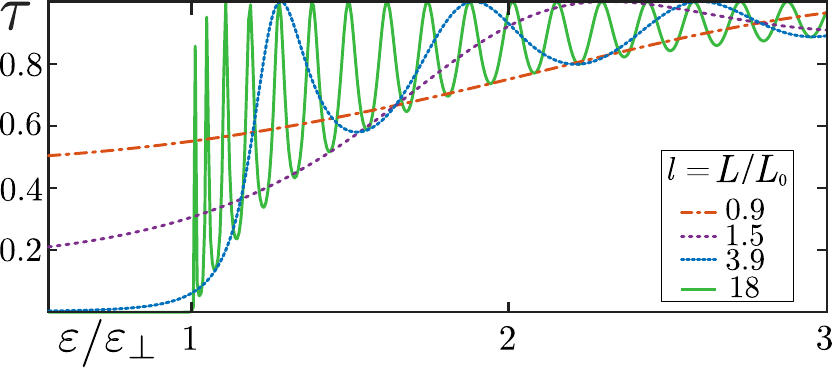}
\end{center}
\caption{Transmission function ${\cal T} \left( \varepsilon \right)$ defined in Eq. \ref{eq:tau}
within a range of lengths $l$ for an island with homogeneous magnetic moment. Energies are
expressed in units of $\varepsilon_{\perp}=J m_{\perp}$, lengths are expressed in units of $L_0=\varepsilon_{\perp} / \hbar v_F$.  }
\label{fig:fig2}
\end{figure}

 In order to calculate the transmission function we proceed as in Ref. \cite{marun}, starting from the evolution operator in space for the whole scattering region. It  reads
 $\hat{\cal U}(x_N,x_0)=\prod_{j=1}^N \hat{\cal U}(x_j,x_{j-1})$,  with
 \begin{eqnarray} \label{eq:evol}
 \hat{\cal U}(x_j,x_{j-1})&=& \exp \{ i \frac{ \varepsilon_{j ||} }{\hbar v_F} L_j  \} \exp\{- i \boldsymbol{\lambda}_j  \cdot \hat{\boldsymbol{\sigma}} \}  \\
 & = & \exp \left\lbrace i \frac{ \varepsilon_{j ||} }{\hbar v_F} L_j  \right\rbrace 
\left[  \hat{\sigma}_0 \cos{\lambda_j} - i {\bf n}_j  \cdot \hat{\boldsymbol{\sigma}}\sin{\lambda_j} \right], \nonumber
 \end{eqnarray}
being $\boldsymbol{\lambda}_j= \left(i\; \varepsilon_{j \perp} \sin \phi_j, -i \;\varepsilon_{j \perp} \cos \phi_j, \varepsilon \right) L_j /(\hbar v_F) $, with $\varepsilon_{||,\perp}= J m_{||,\perp}$,
and  $ {\bf n}_j= \boldsymbol{\lambda}_j/\lambda_j$. The transmission function is the inverse of the element $2,2$ of the transfer matrix, which is, in turn, the inverse of the matrix  $\hat{\cal U}(L,0)$.
Hence,  ${\cal T}(\varepsilon)=| \mbox{Det}[\hat{\cal U}(x_N,x_0)]/ {\cal U}(x_N,x_0)_{1,1}|^2$.

{\em Single homogeneous island}.
We start by discussing the case of an homogeneous domain of length $L$, described by the previous Hamiltonian with a single piece, $N=1$. The resulting transmission function is
\begin{equation} \label{eq:tau}
{\cal T}(\varepsilon)= \frac{|\varepsilon_{\perp}^2- \varepsilon^2|}{|\varepsilon_{\perp}^2- \varepsilon^2|\cos^2\lambda + \varepsilon^2 \sin^2 \lambda},
\end{equation}
being $\lambda=  r  l$, with $l = L/ L_0$, $L_0 = \hbar v_F/\varepsilon_{\perp} $ and  $r=\sqrt{(\varepsilon/\varepsilon_{\perp})^2-1}$.
Notice that the transmission function does not depend on the detailed orientation of the magnetic moment but only on the projection $m_{\perp}$ perpendicular to the direction of the spin-orbit interaction of the material. It is also symmetrical to $\varepsilon = 0$. 
The latter introduces an effective coupling between the two Kramer's partners, that may open a gap in the spectrum of magnitude $\varepsilon_{\perp}$. 
 
The behavior of ${\cal T}(\varepsilon)$ is illustrated in Fig.\ref{fig:fig2}, where we see its dependence on the length of the island.
For short islands, there is a sizable tunneling amplitude through the magnetic island, while as the length of the magnet increases, the transmission function tends to a step function close to $\varepsilon \sim \varepsilon_{\perp}$. 
We get the following behavior of the transmission function at the opening of the gap as a function of length ${\cal T}(\varepsilon_{\perp}) = \left[1+l^2\right]^{-1}$, with $l=L/L_0$,
  while the slope behaves as $d {\cal T}/d\varepsilon|_{\varepsilon_{\perp}} = 2l^4[1+l^2]/3[1+l^2]^3$, which saturates at the value of $ 2/3$, for increasing $l$. 
For energies $\varepsilon > \varepsilon_{\perp}$, ${\cal T}(\varepsilon)$ exhibits oscillations with maxima ${\cal T}^{\rm max}(\varepsilon_n) =1$ and minima
  ${\cal T}^{\rm min}(\varepsilon_m) = 1- \left(\varepsilon_{\perp}/\varepsilon_m\right)^2 $
  at energies satisfying 
 $\left(\varepsilon_{n (m)}\right)^2\!\!\!=\! \left(\varepsilon_{\perp}\right)^2 + \left(\pi \alpha_{n (m) } \hbar v_F/L\right)^2$, with $\alpha_{n (m)}$ being an
 integer (half-integer) number, respectively. 
 
 
 
\begin{figure}
    \includegraphics[width=\columnwidth]{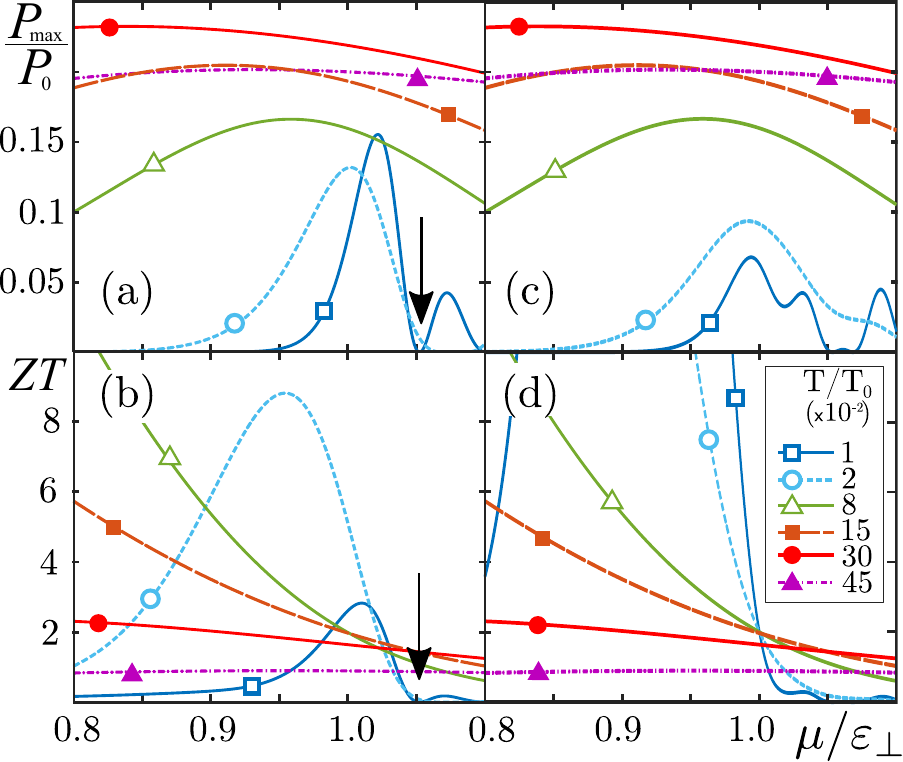}
    \caption{Maximum power (upper panels) and figure of merit $ZT$  (lower panels), for a single magnetic domain of $\textcolor{red}{l} = 10 $ (a)-(b)  and $\textcolor{red}{l}= 20 $ (c)-(d). 
       The maximum values  in (a) and (b) are $P_{\rm max}( T = 0.3 ) = 0.240 P_0$ (a), $ZT(T=0.08 ) = 60$ (b), $P_{\rm max}(T = 0.45) = 0.244 P_0$ (c), and $ZT(T=0.02) = 274$ (d).
   The temperatures are expressed in units of $T_0=\varepsilon_{\perp}/k_B$ and the power is expressed in units of $P_0 = (k_B \Delta T)^2 / h$. Other details are the same as in Fig.\ref{fig:fig2}. }
    \label{fig:fig3}
\end{figure}

The impact of the transmission function on the thermoelectric performance of the heat engine is illustrated in Fig. \ref{fig:fig3} for two lengths of the magnetic domain, in a range of chemical potentials close to the edge of the energy gap, within a temperature range scaled by the reference temperature $T_0=\varepsilon_{\perp}/k_B$. 
For the shortest length shown in panels (a) and (b), $l=10$, ${\cal T}(\varepsilon_{\perp}) <0.01$ and $d {\cal T}/d\varepsilon|_{\varepsilon_{\perp}} \sim 0.65 $, i.e. close to the maximal slope ($2/3$), 
implying a pronounced step in the transmission function at the closing of the energy gap.
The plots shown in Figs.(c) and (d) correspond to a longer island of length $l=20$, for which 
the step function is slightly more pronounced. For very low temperatures, within a scale $k_B T$ smaller than the width of the peaks of ${\cal T}(\varepsilon)$, both $P_{\rm max}$ and $ZT$ vanish
for $\mu=\varepsilon_n$  (see arrows in panels (a) and (b)).  As the temperature increases, the behavior of these quantities is ruled by the effect of several peaks. At sufficiently high temperature, such that  several maxima of ${\cal T}(\varepsilon)$ are included in an energy window of width $k_B T$, the behavior is dominated by the average between the envelopes for the minima and the maxima of ${\cal T}(\varepsilon)$. The resulting function is approximately a smoothed step-function, independently of the length of the island. For this reason, $P_{\rm max}$ shows a wide maximum centered 
at $\sim |\varepsilon_{\perp}- \mu| \sim k_B T$ \cite{whitney1,peter}. The maximum is as high as $\sim 0.244 P_0$, i.e. $ \sim 75 \% $ of the bound $0.32 P_0$. 
${\cal T}(\varepsilon)$ is symmetric with respect to $\varepsilon=0$ and has a well  of unit depth and width $\sim 2 \varepsilon_{\perp}$. This feature dominates the 
 behavior of the power and $ZT$ at high temperatures. These
 properties depend mildly on the length of the island.
Details of the effect of the different features of ${\cal T}(\varepsilon)$ on the thermoelectric response as a function of $T$ are presented in
the supplementary material (SM) \cite{sm}.

\begin{figure}
    \includegraphics[width=\columnwidth]{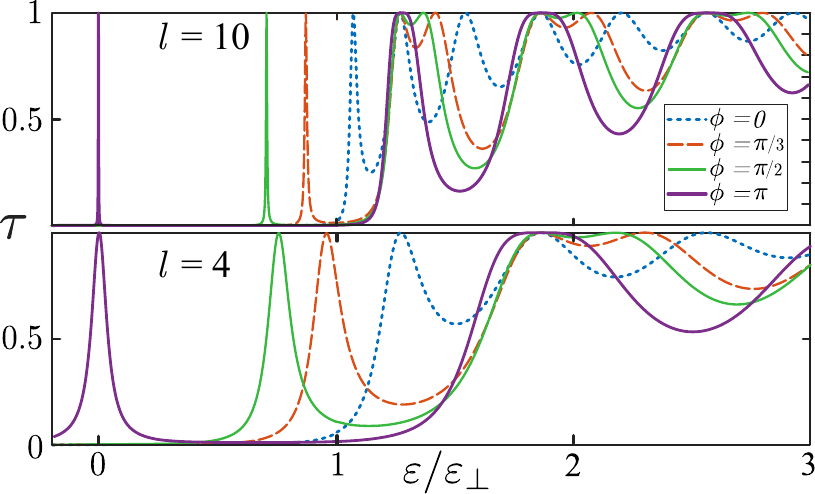}
    \caption{
   Transmission function ${\cal T}(\varepsilon)$ defined in Eq. \ref{eq:tau} for two magnetic domains of equal size ($l = {4,10}$) with the perpendicular component of the magnetic moments oriented with a relative tilt $\phi$.}
    \label{fig:fig4}
\end{figure}

{\em Two domains}.
We now turn to analyze the case where we have two pieces, corresponding to $N=2$ in Eq.(\ref{eq:evol}) with $L_1=L_2=L$, $\phi_1=0$, $\phi_2=\phi$, and $\varepsilon_{\perp,1}=\varepsilon_{\perp,2}=\varepsilon_{\perp}$. The resulting transmission function reads,  
\begin{align}
{\cal T}(\varepsilon) &=\left \{  \left[\cos^2\lambda+\frac{\sin^2 \lambda}{r^2} \left(\cos\phi- \frac{\varepsilon^2}{\varepsilon_{\perp}^2} \right ) \right ]^2 \right. + \nonumber \\ 
& \left. \left[- \frac{\varepsilon}{\varepsilon_{\perp}} \frac{\sin 2\lambda}{r}+\sin\phi \frac{\sin^2\lambda}{r^2} \right ]^2 \right \}^{-1}.
\end{align}
The new feature in the present case, in comparison to the case of a single magnetic moment, is the existence of resonances  within the gap,  $|\varepsilon| < \varepsilon_{\perp}$,
for $\phi \neq 0$. The position of the resonant state depends on the phase difference $\phi$.
For $\phi = \pi$, Eq. (\ref{ham}) coincides in that case with the model introduced by Jackiw and Rebbi \cite{jare,ssh}, which has a topological zero mode localized at the domain wall boundary.
 In Ref. \onlinecite{sm} we analyze the impact of the length of the domains on the 
 width of the resonant state. We also show that  this feature is robust under weak disorder in the length of the domains and the orientation
of the magnetization along each domain.

\begin{figure}
    \centering
    \includegraphics[width=\columnwidth]{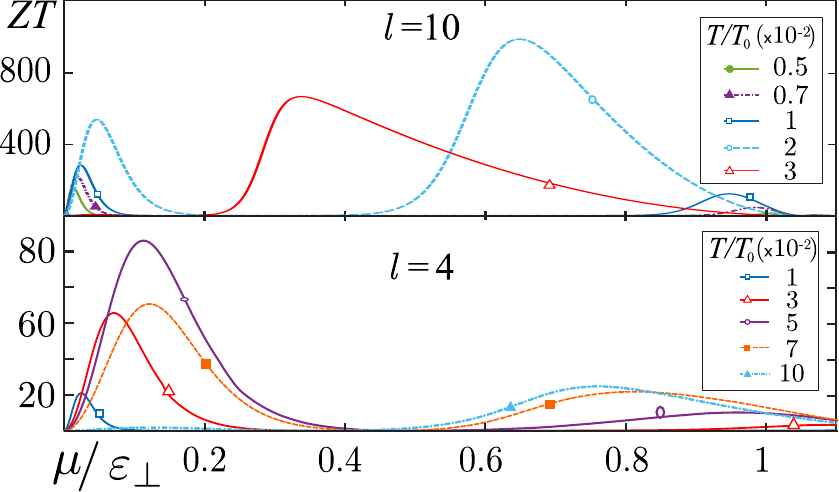}
    \caption{Figure of merit $ZT$ for two magnetic domains of length $l = {4,10}$, with the perpendicular component of the magnetic moments  tilted in  $\phi=\pi$.  
    Other details are similar to previous figures. }
    \label{fig:fig5}
\end{figure}

The behavior of the transmission function for two domains is illustrated in Fig. \ref{fig:fig4} for a set of orientations. 
The upper and lower panels show the transmission function for $l=10$ and  $l=4$ for each domain, respectively. Note that the width of the resonance decreases for increasing $l$.
The corresponding thermoelectric response is shown in Fig. \ref{fig:fig5}. 
Close to the  edge of the gap, the minima  of ${\cal T}(\varepsilon)$ for $\phi=\pi$, are deeper than the ones for a single domain (see Eq. 6 of Ref. \onlinecite{sm}). Notice that the latter corresponds to a single domain with total length $2L$. On the other hand, for $\phi=\pi$,
the energy difference between peaks is twice the one for $\phi=0$.
Hence, for two domains with $\phi=\pi$, the first peak after the closing
of the gap is expected to generate a thermoelectric response with a high figure of merit, similar to that of a Lorenzian function within a range of temperatures larger than in the case of a single ferromagnetic one. 
For $\mu \sim k_B T$ the thermoelectric response is dominated by the resonance within the gap. This leads to  high values of  $ZT$ for $k_B T \lessapprox 10 \gamma$, being $\gamma$ the width of the resonance, which
depends on the domain length. These details are discussed in  Ref. \cite{sm}. For higher temperatures, the transport behavior is dominated by the Heaviside-step function and well-shaped envelopes of the transmission 
 function, and the thermoelectric response is similar to the one discussed for a single domain.

{\em Conclusions.}
We have analyzed the transmission function characterizing the coherent transport of electrons in a structure consistent of a pair of helical edge states of a 2D TI coupled by a magnetic 
island with a magnetic moment having a component perpendicular to the direction of the spin orbit of the TI. We have shown that this setup has the necessary conditions to achieve high
performance thermoelectricity. The key is the opening of a gap in the spectrum of the helical edges with a steep increase of the transmission function at the opening of the propagating modes in the spectrum. Depending on the energy range and the configuration of the magnetic domains, the transmission function has features akin to a theta-function, as well as with features akin to a delta-function, which are known to
be optimal for high-power production and figure of merit, respectively. Due to the resonant states in the gap for two magnetic domains, very large values of the figure of merit, $ZT>100$, are attained for  the heat-engine
and refrigeration modes. Our calculations focus on a single
 pair of edge states, but the currents simply  scale in a factor two when the pair at the opposite edge is also considered. 
The range of operation is set by the magnetic gap $\varepsilon_{\perp}$. 
For a single domain generating an effective magnetic field of $\sim 1.8 - 4$ T \cite{Scheunert2016}, we estimate $\varepsilon_{\perp} \sim 1 - 2 \times 10^{-4}  eV$, corresponding to reference temperatures $T_0 \sim 1.2 - 2.4$ K. According to our study, such a device with a length of the magnetic island of $ \sim 10 L_0$, being $L_0= \varepsilon_{\perp} / \hbar v_F$,
operates as a heat engine at a high performance ($\sim 75\%$ of the
optimal bound)  regarding power generation with a figure of merit $ZT \gg 1$ 
 for $T < 0.5 T_0$. Taking estimates for  the Fermi velocity of the helical edge states in
 quantum wells of HgTe  from Ref. \cite{ti4}, we have $\hbar v_F \sim 0.9 eV/nm$, leading to $L_0 \sim 10- 20 \mu m$.   These parameters are at the state of the art of present experimental realizations. 

{\em Acknowledgements}.
We thank A. Aligia and P. Roura-Bas for carefully reading our manuscript and useful comments.  We acknowledge support from CONICET, Argentina.  We are sponsored by PIP-RD 20141216-4905 
of CONICET,  PICT-2014-2049 and PICT-2017-2726 from Argentina, as
well as  the Alexander von Humboldt
Foundation, Germany (LA).

\newpage

\begin{appendices}

\textbf{SUPLEMENTAL MATERIAL: OPTIMAL THERMOELECTRICITY WITH QUANTUM SPIN-HALL EDGE STATES}

In order to gain insight on the features of the transmission function ${\cal T}(\varepsilon)$ ruling the thermoelectric response of  a single Kramer's pair of helical edge states of the QSH coupled to a nanomagnet, we analyze simpler functions and take them as  reference. In particular, we analyze 
 the thermoelectric response  of
 Heaviside-step (\sref{sec:Heaviside}), well-shaped (\sref{sec:well}) thansmission functions,  and a well-shaped transmission function with a resonance in the energy gap (\sref{sec:well-plus-bound}). In Sec. II we show that at different energy scales, ${\cal T}(\varepsilon)$  for the nanomagnet-QSH system contains ingredients of  these different functions. Depending on the temperature, a particular one becomes dominant. We  show in Sec. III the existence of a resonant state in the case of two magnetic domains with relative orientation $\phi=\pi$.
  Finally, in  \sref{sec:inhom}  we analyze the robustness of the main features of the transmission function against disorder in the orientation of the magnetic moment within the domains.

\begin{figure}[h!]
	\centering
	\includegraphics[width=\columnwidth]{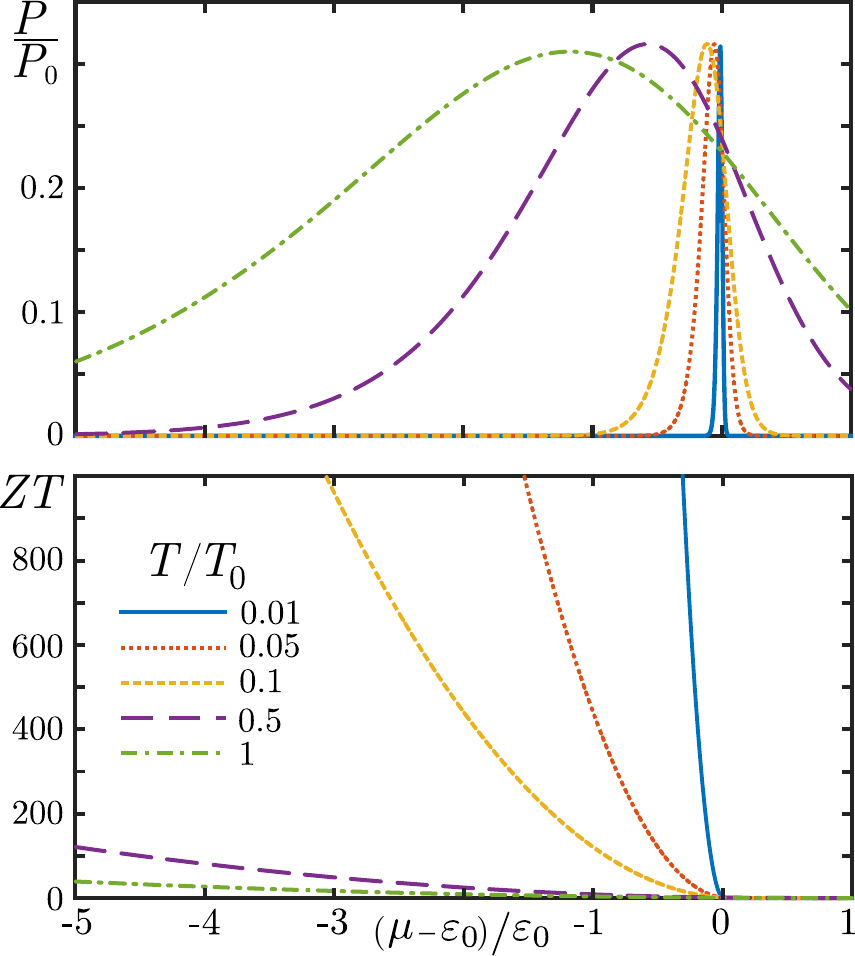}
	\caption{Maximum power $P_{\rm max}$ and $ZT$ for the Heaviside step function defined in Eq. (\ref{step}) for several temperatures, expressed in units of $T_0=\varepsilon_0/k_B$. }
	\label{fig:fig1s}	
\end{figure}

\begin{figure}[h!]
	\centering
	\includegraphics[width=\columnwidth]{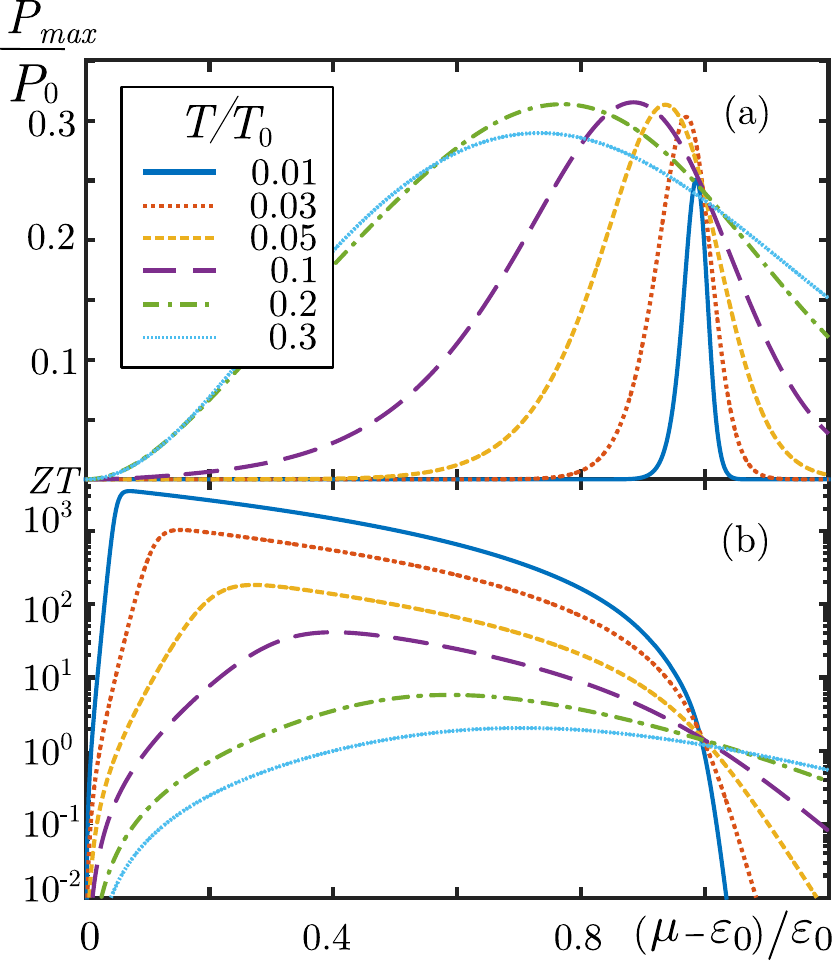}
	\caption{Maximum power and $ZT$ for a well-function defined in Eq. (\ref{well}) as function of $\mu$ for several temperatures. The functions are symmetric with respect to $\mu=0$. Other details are the same as in
	the previous Fig.}
	\label{fig:fig2s}	
\end{figure}

\begin{figure}[h!]
	\centering
	\includegraphics[width=\columnwidth]{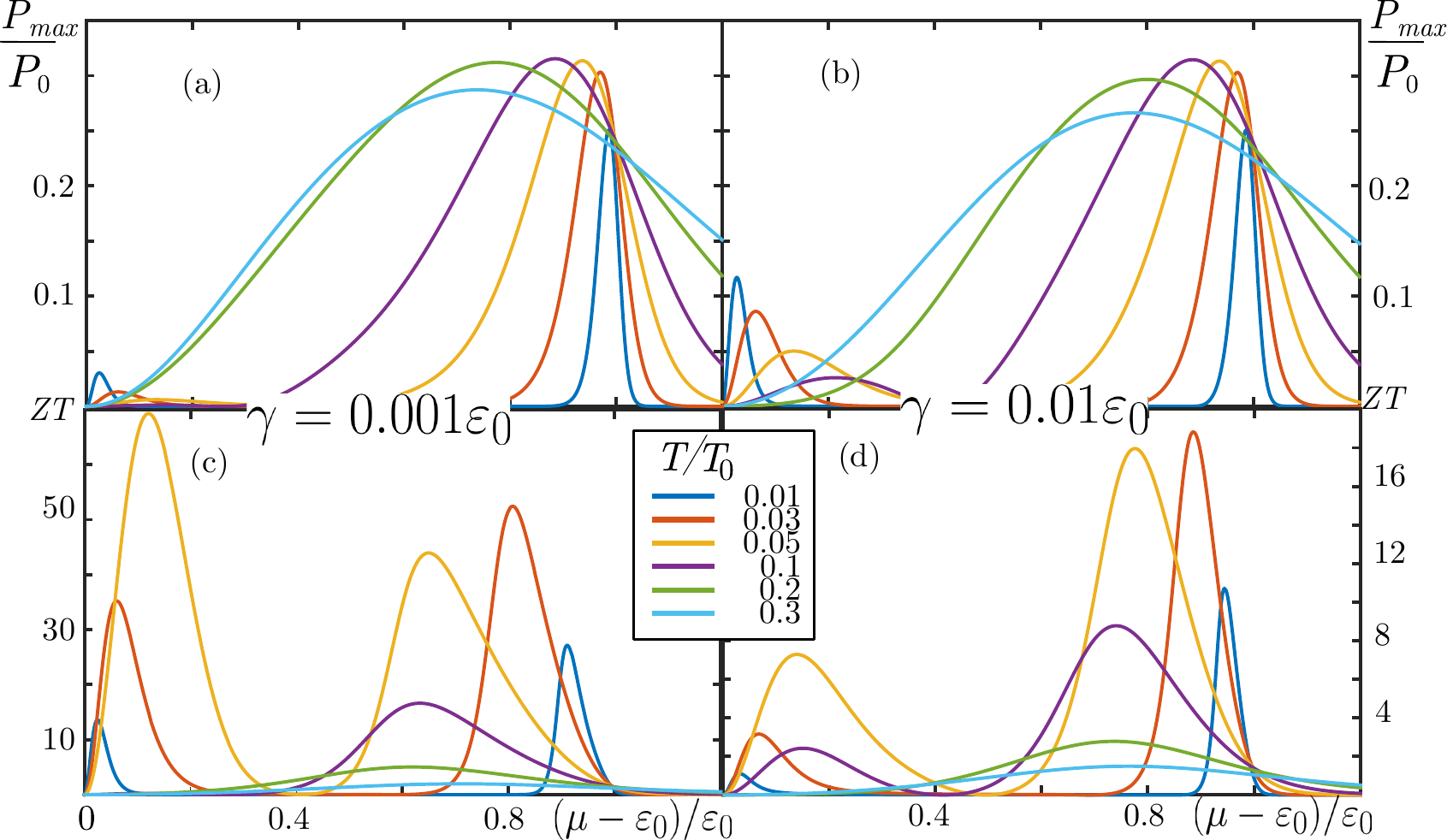}
	\caption{(a-b)Maximum power and (c-d)$ZT$ for the bound state. (a-c) correspond to $\gamma = 0.001 \varepsilon_0$ while (b-d) to $\gamma=0.1\varepsilon_0$. Other details are the same as in the previous figures. }
	\label{fig:fig3s}	
\end{figure}

\subsection{Heaviside-function}\label{sec:Heaviside}
We review the behavior of the figure of merit $ZT$ and the maximum power $P_{\rm max}$, defined, respectively,  in Eqs. (3) and (4) of the main text, for the case of a transmission function with the form of a Heaviside function,
\begin{equation}\label{step}
{\cal T}(\varepsilon)= \theta(\varepsilon-\varepsilon_0).
\end{equation}
  Results are shown in \fref{fig:fig1s}. The upper bound for the maximum  achievable power, $0.32P_0$, where $P_0 = (k_B \Delta T)^2/h$,
is reached \cite{benenti,whitney1} at $\mu-\varepsilon_0 \sim -k_B T$. We see that  $ZT$ achieves high values at low temperatures. However, it is a decreasing function of the temperature, and drops very rapidly to $ZT \sim 0$ in the region of $\mu-\varepsilon_0>0$,
where  ${\cal T} = 1$.

\subsection{Well-shaped function}\label{sec:well}
In this section we consider a well-shaped  function  of the form 
\begin{equation}\label{well}
{\cal T}(\varepsilon) = \theta(\varepsilon-\varepsilon_0) + \theta(-\varepsilon-\varepsilon_0).
\end{equation}
The range $|\varepsilon| < \varepsilon_0$ behaves like   the transmission function of the nanomagnet coupled to the helical edge states within the energy gap for sufficiently large magnets such that $L \gg L_0$. 
In the range $|\varepsilon| > \varepsilon_0$, it is similar to the envelope for the sequence of maxima of the transmission function of the nanomagnet.  Both the maximum power and $ZT$ are even functions of $\mu$. 
Results are shown in \fref{fig:fig2s}
for $\mu>0$.
 For low temperatures, $T<0.2 T_0$, being $T_0=\varepsilon_0/k_B$, the behavior of both, the maximum power and $ZT$ is similar to that of the two Heaviside-type transmission functions. In fact, we can identify in Fig. \ref{fig:fig2s} the same features that we have already analyzed in Fig. \ref{fig:fig1s}. In particular, we see that the maximum power achieves the optimal bound $0.32 P_0$. However, for $T>0.2T_0$, the maximum value becomes a decreasing function of 
 the temperature. The behavior of $ZT$ is similar to that of the step function within the whole range of temperatures.

\subsection{Well-shaped transmission function with a resonant peak}\label{sec:well-plus-bound}
We now consider the thermoelectric response of a transmission function, which consists of  a well-shaped transmission function with a  Lorenzian function  of width $\gamma$ in the center of the well, 
\begin{equation}
{\cal T}(\varepsilon)= \theta(\varepsilon-\varepsilon_0) + \theta(-\varepsilon-\varepsilon_0) + \frac{ \gamma^2}{(\gamma^2+\varepsilon^2)}.
\end{equation}
We focus on $\gamma \ll \varepsilon_0$, which is relevant for the description of a transmission function of the nanomagnet with two magnetic domains, which hosts a narrow resonance inside the gap.
 The corresponding  thermoelectric response is presented in \fref{fig:fig3s} for two different widths of the resonance, $\gamma = 0.001 \varepsilon_0$ and $\gamma = 0.01\varepsilon_0$.
 We start by analyzing the behavior of the maximum power, which is shown in the upper panels of the Fig. 
 By comparing these plots   with those shown in the upper panels of \fref{fig:fig2s} and \fref{fig:fig1s}, we see that the dominant response at all temperatures is due to the
 well-function feature. For low temperatures $T<10 \gamma/k_B$, we can also identify features originated in the response due to the resonant peak. Concretely, we can identify maxima of $P_{\rm max}$
 at $\mu \sim k_B T$. The maxima are anyway much smaller than the optimal bound ($\sim P_0/3$ for the case of $T=0.01T_0$ and $\gamma=0.01$). In the behavior of $ZT$, we can also identify features originated in the resonance and
 on the step function. For $T\leq 0.05 T_0$, the resonance leads to maxima in $ZT$ at values of the chemical potential that are $|\mu| \propto k_B T$, while the step function leads to maxima of $ZT$ at values of the chemical potential
 satisfying $|\mu-\varepsilon_0| \propto k_B T$. Hence, within this low-temperature range, there are two maxima, which become closer one another as the temperature increases. For higher temperature the positions of the two maxima change mildly 
 but the values of $ZT$ decrease rapidly with the temperature.

\section{Thermoelectric regimes of the  nanomagnet coupled to the helical edge states}
We expect different thermoelectric regimes as a function of temperature. The transmission function of the nanomagnet caupled to the edge states  presents ingredients of the three functions described in the previous section. Concretely, we expect that at low temperatures, where resonances and peaks are resolved, the thermoelectric response resembles that of the Lorenzian function, at higher temperatures, the behavior of the Heaviside function becomes dominant and at even higher temperatures, the well-shape due to the gap causes the  decreasing behavior of both $P_{\rm max}$ and $ZT$ as functions of $T$ for all values of the chemical potential $\mu$.

\begin{figure}
	\centering
	\includegraphics[width=\columnwidth]{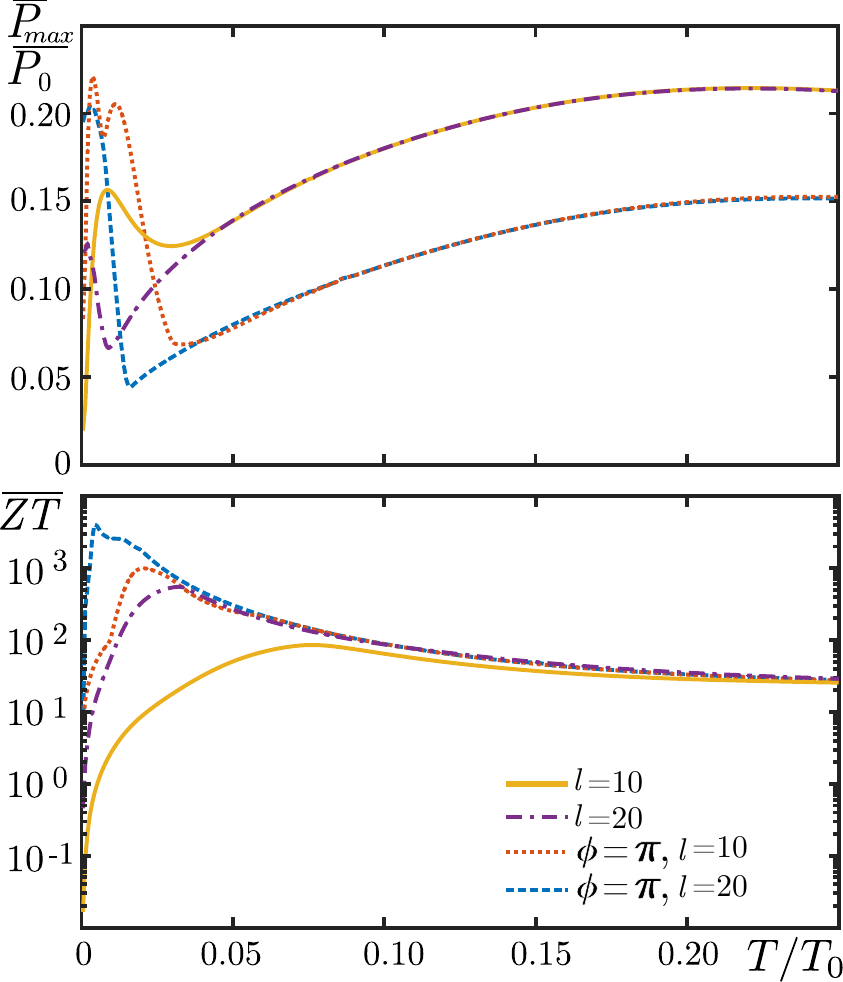}
	\caption{$\overline{ZT}=\mbox{Max}_{\mu} \left[ZT \right]$ and $\overline{P}_{\rm max}=\mbox{Max}_{\mu}\left[P_{\rm max}\right]$, corresponding to the maximum values of $P_{\rm max}$ and $ZT$ over the whole  range of  $\mu$, as functions  of the temperature $T$. AF denotes islands with two domains antiferromagnetically aligned.}
	\label{fig:fig4s}	
\end{figure}

An overview of the different regimes is presented in Fig. \ref{fig:fig4s}. Here, we show the functions $\overline{ZT}=\mbox{Max}_{\mu} \left[ZT \right]$ and $\overline{P}_{\rm max}=\mbox{Max}_{\mu}\left[P_{\rm max}\right]$, where  $\mbox{Max}_{\mu}$ denotes the maximum value of the quantity  over the whole  range of values of $\mu$. The boundary for the lowest-temperature regime is identified with the 
range of $T$ below the one corresponding  to the minimum of $\left[P_{\rm max}\right]$.   This regime is akin to the response due to a Lorenzian-type transmission function and  is 
non-universal, since it depends on the details of the resonant peaks of ${\cal T}(\varepsilon)$. The latter are determined by the length of the island and the number of domains. In the case of two  domains woth $\phi=\pi$, it is possible to distinguish a narrow feature, which is associated to a resonance within the gap, followed by a second feature, associated to the the first peak after the closing of the gap. For islands with a single domain, we can distinguish only one feature, which is associated to the first peak after the closing of the gap. For  larger temperatures, we enter the regime dominated by the Heaviside-type function corresponding to averaging the envelopes of the minima and maxima of the transmission function for $\varepsilon>\varepsilon_{\perp}$. While the maxima corresponds to ${\cal T}=1$, the minima are deeper for two domains with $\phi=\pi$ than for a single domain. Hence, the values of 
$\mbox{Max}_{\mu} \left[P_{\rm max}\right]$ within this regime are higher for a single domain than for two antiferromagnetic ones. For both orientations, the behavior is universal, namely, it does not depend on the length of the island. The figure \ref{fig:fig4s} shows the onset of the third (high-temperature) regime, where the well-shape becomes dominant and $\mbox{Max}_{\mu} \left[P_{\rm max}\right]$ turns to be a decreasing function of $T$. 

\section{Resonant state in the $\phi=\pi$ configuration}\label{sec:resonant}
The existence of a resonant state in the gap for this configuration can be understood after noticing that the Hamiltonian of Eq. (5) of the main text reduces to a 1D Dirac Hamiltonian with a mass, determined by the coupling to $m_{\perp}$,
which changes sign at the boundary between the two domains. This is, precisely the model analyzed by Jackiw and Rebbi \cite{jare}, which is the continuous version of the Su-Schrieffer-Heeger model
\cite{ssh} for polyacetylene. These models are known to host topological zero modes localized at the domain wall.  We now analyze the degree of localization of the zero mode for the case of interest, where the length of the magnets 
(equivalent to the massive region of the Dirac Hamiltonians) is finite.

The inverse of  transmission function of two magnetic domains with length $L$ and a relative orientation $\phi$  of the component of the magnetization perpendicular to the spin-orbit interaction of the TI,  takes the following simple form
\noindent
\begin{align}\label{eq:resonant}
&\left[{\cal T} (\varepsilon,l,\phi)\right]^{-1} = \nonumber \\
& \left [ \cos^2\lambda+\frac{\sin^2\lambda}{r^2}(\cos\phi-x^2) \right ]^2 + 
\left [ -x \frac{\sin 2\lambda}{r}+\sin\phi\frac{\sin^2\lambda}{r^2} \right ]^2, 
\end{align}
\noindent
where $l=L/L_0$, $x = \varepsilon/\varepsilon_\perp$, $r = \sqrt{x^2-1}$ and $\lambda = l r$. Notice  that for $\varepsilon =0$, the latter function reads
\noindent
\begin{align}\label{eq:resonant}
&\!\left[{\cal T} (0,l,\phi)\right]^{-1}\!= \nonumber\\
&\! 
\left[\cos^2(li)+\left(\frac{\sin(li)}{i} \right )^2 \cos\phi \right ]^2 \! +\! \left(\sin\phi \left (\frac{\sin li}{i}  \right )^2 \right )^2
 \nonumber \\ 
&= \left(\cosh^2l+\sinh^2l \cos\phi \right )^2+\sin^2\phi \sinh^2l \nonumber \\
&= \left(1+\left(1+\cos\phi \right )\sinh^2l \right )^2 + \sin^2\phi \sinh^2l, 
\end{align}
\noindent
where we see that  ${\cal T}(0,l,\phi) \sim 0$ for $l>1$ except for  
 $\phi = (2n+1) \pi$,  with $n$ integer, in which case ${\cal T}(0,l,(2n+1)\pi) = 1$.  Therefore, we conclude that for $\phi=(2n+1) \pi$, there is a resonant state in the center of the gap, with energy $\varepsilon=0$.
The width of this resonant state  depends on the length as $e^{-l}$, which means that the width of the resonance decreases with $l$. 

For this configuration there is a simple expression for the minima of the oscillations above the gap. It reads
\begin{equation}
{\cal T}^{min}(\varepsilon_{m})=\left[\left(\varepsilon_m^2-\varepsilon_\perp^2 \right )/\left(\varepsilon_m^2+\varepsilon_\perp^2 \right ) \right ]^2,
\end{equation}
 where  
$\left(\varepsilon_{m}\right)^2\!\!\!=\! \left(\varepsilon_{\perp}\right)^2 + \left(\pi \alpha_{m} \hbar v_F/L\right)^2$, with $\alpha_{m}$ being a
 half-integer number. 

In addition, we  estimate the critical length of the domains for the resonant peak to develop. We use the following criterion to determine the critical length $l_c$  for which the width of the resonant state is smaller than the energy gap,
${\cal T}(\varepsilon/\varepsilon_\perp=0.5)\leq 0.5$, leading to
\begin{equation}
\!\!\!{\cal T}\left ( \frac{\varepsilon}{\varepsilon_\perp}=0.5,l_c \right )=\frac{9/8}{\frac{15}{8}\!\!-\!\!\cosh\left (\sqrt{3}l_c \right )\!\!+\!\!\frac{1}{4}\cosh\left (2\sqrt{3}l_c\right )} .
\label{eq:taul}
\end{equation}
We get $l_c \simeq 0.9 $. Hence, we identify a resonance in the configuration of $\phi=\pi$ for $l>l_c$. We have verified that for angles $\phi\neq \pi$ the criterion does not change significantly and the condition $l > 1$ is enough 
to clearly resolve a resonant state within the gap.

The previous analysis was based on the assumption that the two domains have the same length. In what follows, we analyze the case of magnetic domains with different lengths, focusing on the situation where one of the domains has $l_1>1$ and the second
domain has $l_2 \leq l_1$. The generalization of Eq. (9) of the main text to the present case reads
\begin{multline}
\left [{\cal T}(\lambda_1,\lambda_2,\phi,x)  \right ]^{-1}  =  \\
\left[\cos\lambda_1\cos\lambda_2+\frac{\sin\lambda_1\sin\lambda_2}{r^2}\left(\cos\phi-x^2 \right ) \right ]^2 \\ 
+ \left[ \sin\phi \frac{\sin\lambda_1\sin\lambda_2}{r^2}-\sin(\lambda_1+\lambda_2)\frac{x}{r} \right]^2.
\end{multline}
For $\phi=\pi$, the height of the resonant level is given by 
\begin{equation}\label{tau12}
{\cal T}(\lambda_1,\lambda_2,\phi=\pi,x=0) = \frac{1}{\cosh\left(l_1-l_2\right)}.
\end{equation}
We see that the transmission function is $\tau=1$ for $l_1=l_2$, consistently with our previous analysis of two equaly sized domains. For $l_1 \neq l_2$, $\tau$ drecreases and becomes $\tau\sim 0$ if $l_2\ll l_1$ (recall that $l_1 >1$).  
From Eq. (\ref{tau12}) we find that for $|l_1-l_2|\leq 0.5$ and $l_1>1$, $\tau$ hosts a clear resonance with $\tau(0)>0.9$.
 \begin{figure}
	\centering
	\includegraphics[width=\columnwidth]{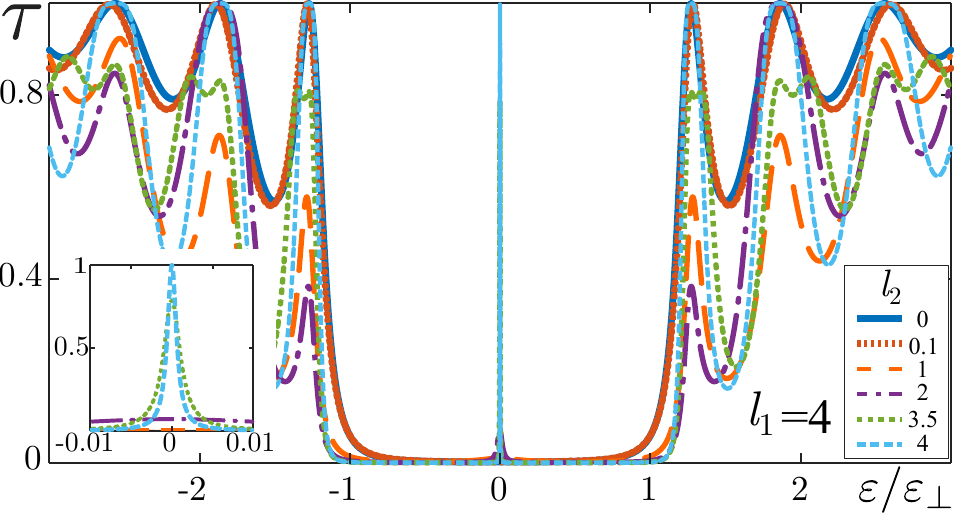}
	\caption{Transmission function for a two-domain configuration with $\phi=\pi$. The first island has a fixed length and we consider several lengths of the second island. The inset shows a detail of the bound state at zero energy}
	\label{fig:fig5s}	
\end{figure}
 The behavior of $\tau(\varepsilon)$ for $l_1 \neq l_2$ is illustrated in \fref{fig:fig5s} for $l_1=4$ and $l_2<l_1$. We see that for $l_1-l_2 > 2$, the resonance is no longer distinguished within the gap. However, for small difference in the length of the two magnetic domains, not only the resonant peak, but also the behavior of $\tau(\varepsilon)$ above the gap is practically unaffected.

\section{Robustness of the features of the transmission function against inhomogeneities in the orientation of the magnetic moment}\label{sec:inhom}
We now turn to analyze the effect of an inhomogeneity in the orientation of the magnetic moment within each domain. To this end, we divide each magnetic domain in $n$  pieces of the same length, along which the phase gets a random component. Hence, the orientation of the magnetic moment within each of these pieces is $\phi_j = \phi^0+\delta\phi_j, \; j=1, \ldots, n$,  where $\delta \phi_j$ is  a random component of the phase within the $j-$th subdomain, while $\phi^0 = 0,\pi$ for the first and second domain, respectively. According to the previous analysis, the partitions must satisfy $l/n <<1$, in order to be considered as a perturbation over the main magnetic
 configuration of the domain. In fact, for $l/n \sim 1$, each of these partitions would separately open a gap and would behave as an additional magnetic domain.

\begin{figure}
	\centering
	\includegraphics[width=\columnwidth]{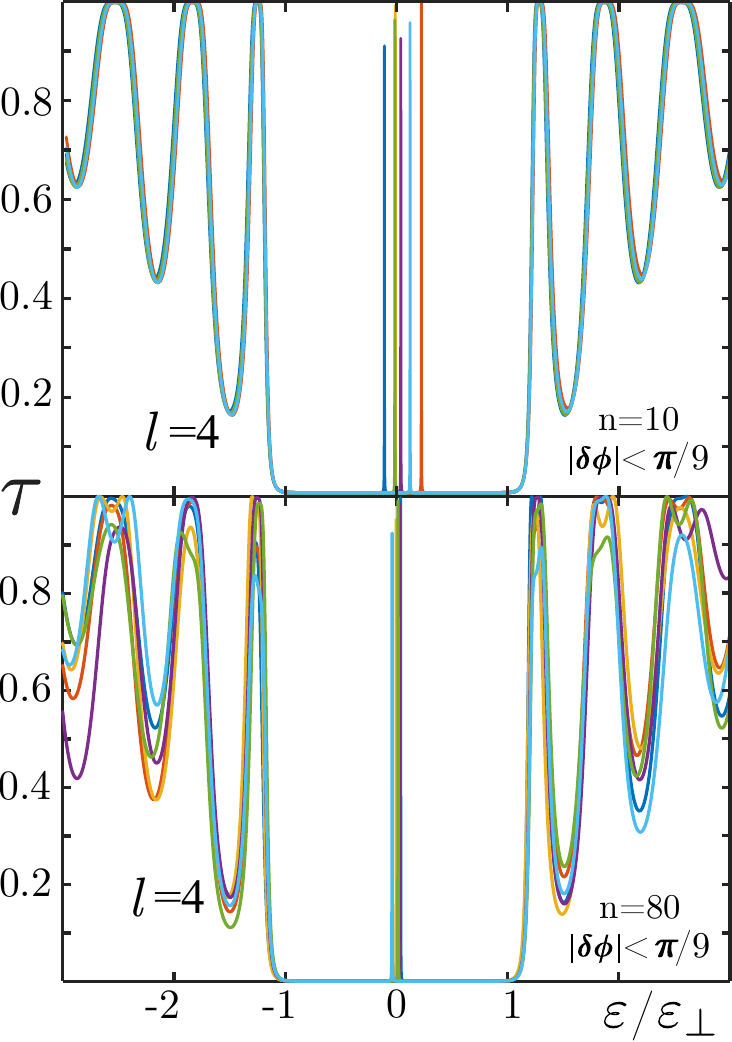}
	\caption{Transmission function for two domains of equal length $l=4 $ with orientation of the magnetic moments  $\phi^0=0,\pi$ and random piece-wise random fluctuations $\delta \phi_j=\pm \pi/9,\;j=1,\ldots,n$, within
	 within  $n=10$ partitions of equal length (top) and $n=80$ (bottom). Different colors correspond to different realizations of disorder.}
	\label{fig:fig6s}	
\end{figure}

\begin{figure}
	\centering
	\includegraphics[width=\columnwidth]{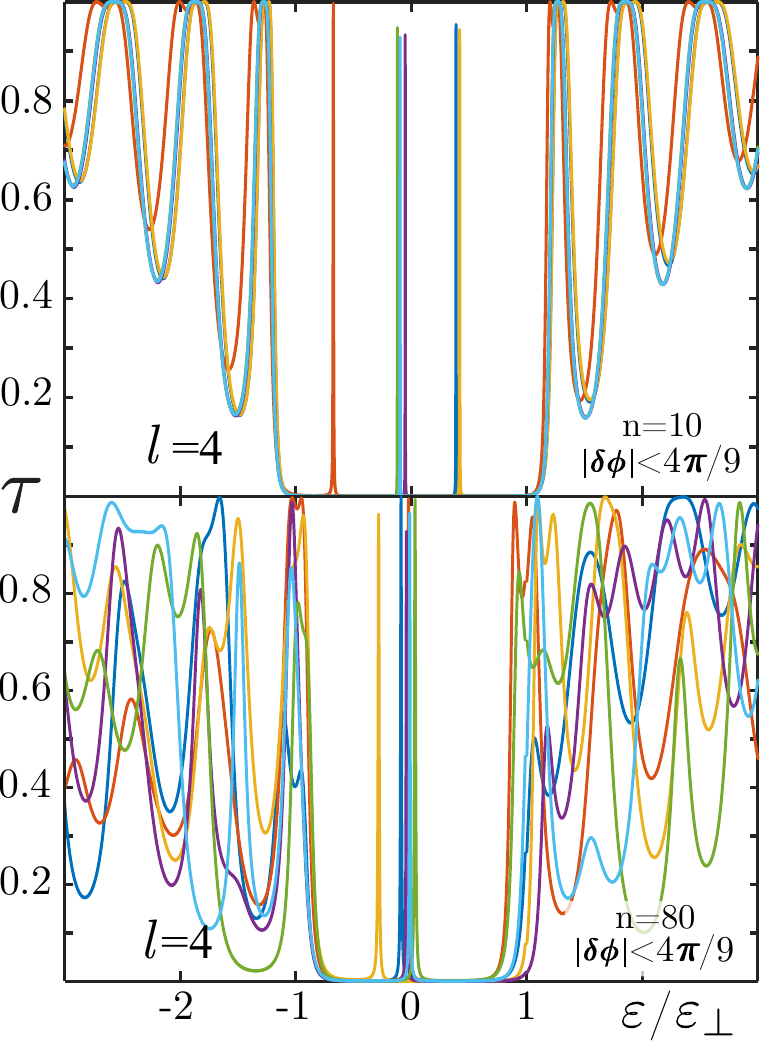}
	\caption{Same as Fig. \ref{fig:fig6s} with $\delta \phi_j=\pm 4 \pi/9$. }
	\label{fig:fig7s}	
\end{figure}

Examples are shown in Figs. \ref{fig:fig6s} and \ref{fig:fig7s} for weak and  strong amplitude in the random component of the magnetic moment, respectively. In each case, we compare the behavior of different numbers of partitions $n$.
In the case of weak disorder shown in Fig. \ref{fig:fig6s}, we see that the behavior of the transmission function above the gap is almost unaffected by the inhomogeneous orientation of the magnetic moment, while the position of the resonant peak 
is slightly shifted away from $\varepsilon=0$. Albeit, the width of the latter remains unaffected. The shift becomes smaller as  the number of partitions increases, as seen in the comparison between the top and bottom panels. 
The case of strong fluctuations ($\delta \phi_j =\pm 4\pi/9$) is analyzed in Fig. \ref{fig:fig7s}. The behavior of the resonance is similar to the case of weak disorder. For increasing fluctuations of the magnetic moment, the shift of the resonance
away from $\varepsilon=0$ is larger. For small number of partitions $n$ (see top panel) the pattern  of maxima and minima above the gap becomes also affected. However,  the overall structure of $\tau(\varepsilon)$, including the existence of  a resonant peak, the clear opening of the gap and a series of peaks with an envelope defined by a Heaviside function are preserved. For larger number of partitions (see bottom panel), all the features, including the resonant peak, as well as the pattern
of maxima and minima above the gap are mildly affected. 

The analysis of this and the previous sections lead us to conclude that the performance of the setup is very robust under  weak fluctuations in the orientation of the magnetic moment, as well as under fluctuations in the length of the two domains.

\end{appendices}

\end{document}